\documentclass[aps,pra,twocolumn]{revtex4}
\RequirePackage{latexsym}

\def\op#1{\hat{#1}}
\def\ket#1{| #1 \rangle}
\def\bra#1{\langle #1 |}

\def\vec#1{{\bf #1}}

\def\trace#1{\mathop{\rm Tr}\nolimits ( #1)} 

\newtheorem{definition}{Definition}
\newtheorem{corollary}{Corollary}
\newtheorem{lemma}{Lemma}
\newtheorem{theorem}{Theorem}
\newtheorem{example}{Example}
\newenvironment{proof}{\par {\sc Proof:}}{$\Box$\par}
\begin{document}
\bibliographystyle{prsty}
\title{Complete controllability of quantum systems}
\author{S.~G.\ Schirmer, H.\ Fu and A.~I.\ Solomon}
\affiliation{Quantum Processes Group, The Open University, Milton Keynes, MK7 6AA, United Kingdom}
\date{November 25, 2000} 
\begin{abstract}
Sufficient conditions for complete controllability of N-level quantum systems 
subject to a single control pulse that addresses multiple allowed transitions 
concurrently are established.  The results are applied in particular to Morse 
and harmonic oscillator systems, as well as some systems with degenerate energy 
levels.  Controllability of these model systems is of special interest since 
they have many applications in physics, e.g., Morse and harmonic oscillators
serve as models for molecular bonds, and the standard control approach of 
using a sequence of frequency-selective pulses to address a single transition 
at a time is either not applicable or only of limited utility for such systems.
\end{abstract}
%Insert valid PACS numbers
\maketitle
\section{Introduction}
\label{sec:Intro}
Recent advances in laser technology have opened up new possibilities for laser 
control of quantum phenomena such as control of molecular quantum states, chemical
reaction dynamics or quantum computers.  This has prompted researchers to study
these systems from a control-theoretical point of view, in particular in view of
the limited success of initially advocated control schemes based largely on 
physical intuition in both theory and experiment \cite{SCI288p824}.

One issue that arises is the question whether, or under which conditions, it is 
possible to control a quantum system in such a way as to achieve any physically 
permitted evolution of the system.  Complete controllability is an important 
theoretical concept that also has significant practical implications.  For example,
it has been shown that kinematical constraints on the evolution of non-dissipative 
quantum systems give rise to universal bounds on the optimization of observables 
\cite{PRA58p2684} and that the practical issue of dynamical realizability of these
bounds depends on the controllability of the system \cite{00SL}.  Controllability 
is also important in quantum computation as it is directly related to the question
of universality of a quantum computation element \cite{PRA54p1715}.

It has been shown that an atomic system with $N$ accessible energy levels, which 
are sufficiently separated to allow control based on frequency discrimination, is 
completely controllable using a sequence of frequency-selective pulses that address
only a single transition at a time \cite{PRA61n032106}.  Although this approach is
enormously useful, control based on frequency discrimination is not applicable to 
systems with equally spaced energy levels such as truncated harmonic oscillators, 
and problematic for systems with almost equally spaced energy levels such as Morse
oscillators, since in this case any external field is likely to address multiple 
transitions simultaneously.  Moreover, systems with degenerate energy levels also 
present problems for this technique.

In this paper we therefore concentrate on complete controllability of quantum 
systems subject to a single control field, e.g., a laser pulse or magnetic field, 
that addresses multiple transitions concurrently.  In particular, we study the 
problem of controllability of a system subject to a single control, for which the 
interaction with the control field is determined by the dipole approximation.  The 
systems considered are assumed to be non-decomposable, i.e., systems that can be 
decomposed into non-interacting subsystems are excluded.  Note that such systems
can never be completely controllable \cite{00SL}.  

It will be shown in particular that a non-decomposable quantum system with dipole
interaction is completely controllable with a single control pulse if there is some
anharmonicity in the energy levels.  It must be noted however that it is sufficient
if, e.g., the transition frequency for the first transition is different from all 
the other transition frequencies.  This is a much weaker condition than frequency 
discrimination, which requires that all the transition frequencies are sufficiently
different.  Furthermore, for a system with no anharmonicity, i.e., equally spaced
energy levels, we demonstrate that complete controllability depends on the values
of the transition dipole moments of the system, and establish sufficient criteria 
for complete controllability.  The controllability of some systems with degenerate
energy levels is also discussed and examples of systems that are not completely 
controllable are presented.
\section{Quantum Control System}
\label{sec:model}
Given any $N$-level quantum system, the Hamiltonian of the unperturbed system can 
be written as
\begin{equation} \label{eq:Hzero}
  \op{H}_0 = \sum_{n=1}^N E_n \ket{n}\bra{n},
\end{equation}
where $\{\ket{n}: n=1,\cdots,N\}$ is a complete set of orthonormal eigenstates and 
$E_n$ are the corresponding energy levels of the system.  

The application of external control fields perturbs the system and gives rise to a 
new Hamiltonian $\op{H}=\op{H}_0+\op{H}_I$, where $\op{H}_I$ is an interaction term.
In the control--linear approximation the interaction term is of the form
\begin{equation} \label{eq:HI}
  \op{H}_I = \sum_{m=1}^M f_m(t) \op{H}_m,
\end{equation}
where $f_m(t)$ for $m=1,\cdots,M$ are independent control fields, and the operator 
$\op{H}_m$ represents the interaction of the field $f_m(t)$ with the system.  The 
off-diagonal elements of $\op{H}_m$ depend on the transition dipole moments $d_{n,n'}$
for transitions between energy eigenstates.  Each $\op{H}_m$ is Hermitian since the 
transition dipole moments satisfy $d_{n',n}=d_{n,n'}^*$, where $d_{n,n'}^*$ is the 
complex conjugate of $d_{n,n'}$.  In this paper we are particularly interested in 
the case $M=1$, i.e., a single control pulse, for which the interaction operator is
of dipole form
\begin{equation} \label{eq:Hone}
   \op{H}_1=\sum_{n=1}^{N-1} d_n (\ket{n}\bra{n+1}+\ket{n+1}\bra{n}), \quad d_n\neq 0.
\end{equation}
Note that we exclude systems for which any of the transition dipole moments $d_n\equiv 
d_{n,n+1}=d_{n+1,n}$ vanish since these systems can be decomposed into non-interacting 
subsystems.

An arbitrary initial state of the system can be represented by a density matrix 
$\op{\rho}_0$ that evolves according to the dynamical law
\begin{equation} \label{eq:dynlaw}
  \op{\rho}(t) = \op{U}(t,t_0) \op{\rho}_0 \op{U}(t,t_0)^\dagger,
\end{equation}
where $\op{U}(t,t_0)$ is the time-evolution operator, which satisfies the 
Schr\"odinger equation
\begin{equation} \label{eq:schrodinger}
  i\hbar \frac{\partial}{\partial t} \op{U}(t,t_0)
 = (\op{H}_0 + \op{H}_I) \op{U}(t,t_0)
\end{equation}
with initial condition $\op{U}(t_0,t_0)=\op{1}$.  

We say the system is initially in a pure state if $\mbox{Tr}(\rho_0^2)=1$.  In 
this case the initial state of the system can also be represented by a normalized 
wavefunction $\ket{\psi_0}$, which is either an energy eigenstate $\ket{n}$ or a 
superposition of energy eigenstates 
\begin{equation}
  \ket{\psi_0} = \sum_{n=1}^N c_n \ket{n},
\end{equation}
where the $c_n$ are complex coefficients that satisfy the normalization condition
$\sum_n c_n c_n^*=1$ \cite{95Mukamel}.  The time evolution of a pure state 
represented by a wavefunction $\ket{\psi(t)}$ is 
\begin{equation}
   \ket{\psi(t)} = \op{U}(t,t_0) \ket{\psi_0}
\end{equation}
where $\ket{\psi_0}=\ket{\psi(t_0)}$ and $\op{U}(t,t_0)$ is the time-evolution 
operator as defined above.
\section{Criteria for Complete Controllability}
\label{sec:control}
Since $\op{H}_0+\op{H}_I$ is Hermitian, (\ref{eq:schrodinger}) implies that the time 
evolution operator $\op{U}(t,t_0)$ is unitary.  Hence, examination of (\ref{eq:dynlaw})
reveals that only target states $\op{\rho}(t_F)$ that are related to the initial state
$\op{\rho}_0$ by $\op{\rho}(t_F)=\op{U}\op{\rho}_0\op{U}^\dagger$, where $\op{U}$ is a
unitary operator, are kinematically admissible.  However, in general, not all of these
states can actually be dynamically reached, unless the dynamical Lie group generated 
by $i\op{H}_0$ and $i\op{H}_m$, $m=1,2,\cdots,M$, is the unitary group $U(N)$.  (See 
appendix \ref{appendix:a} for a discussion of this requirement.)  This motivates the 
\begin{definition} \label{def:one}
A quantum system $\op{H}=\op{H}_0+\op{H}_I$ with $\op{H}_0$ and $\op{H}_I$ as in 
(\ref{eq:Hzero}) and (\ref{eq:HI}) is \emph{completely controllable} if every unitary 
operator $\op{U}$ is accessible from the identity operator $\op{1}$ via a path
$\gamma(t)=\op{U}(t,t_0)$ that satisfies (\ref{eq:schrodinger}).
\end{definition}

Complete controllability implies that any kinematically admissible target state can 
be dynamically reached from the initial state by driving the system with a suitable 
control field.  If the system is initially in a mixed state represented by a density
matrix $\op{\rho}_0$ then this means that any other kinematically admissible mixed 
state can be dynamically reached.  Similarly, if the system is initially in a pure 
state represented by a normalized wavefunction $\ket{\psi_0}$ then complete 
controllability guarantees that every other pure state represented by a normalized 
wavefunction $\ket{\psi_1}$ can be dynamically reached from the initial state.  In
\cite{00SL} it is furthermore shown that complete controllability implies dynamical
realizability of the universal kinematical bounds on the optimization of observables
for (non-dissipative) quantum systems.

It is apparent that if the dimension of the Lie algebra $L_0$ generated by the 
operators $\{\op{H}_0,\cdots,\op{H}_M\}$, or more accurately, their skew-Hermitian 
counterparts $\{i\op{H}_0,\cdots,i\op{H}_M\}$ is $N^2$ then $L_0$ is the Lie algebra
of skew-Hermitian $N\times N$ matrices $u(N)$.  Ramakrishna \emph{et al} have shown 
in \cite{PRA51p960}, using results by Jurdjevic and Sussmann \cite{72JS}, that in 
this case the dynamical Lie group of the system is the unitary group $U(N)$.  Noting
that the dimension of $u(N)$ is $N^2$ and that any Lie algebra of skew-Hermitian 
$N\times N$ matrices of dimension $N^2$ is (isomorphic to) $u(N)$ we have therefore
\begin{theorem} [Ramakrishna et al] \label{thm:cond}
A necessary and sufficient condition for complete controllability of a quantum system 
$\op{H}=\op{H}_0+\op{H}_I$ with $\op{H}_0$ and $\op{H}_I$ as in (\ref{eq:Hzero}) and 
(\ref{eq:HI}) is that the Lie algebra $L_0$ has dimension $N^2$.
\end{theorem}
This theorem provides a condition for complete controllability of a quantum system 
that can easily be verified by computing the Lie algebra generated by $\op{H}_0,\cdots,
\op{H}_M$ and determining its dimension.  

A basic algorithm for constructing a basis for the Lie algebra $L_0$ in terms of 
iterated commutators is presented in Table \ref{table:algo}.  It can be optimized to
increase the speed of the computation and to improve the accuracy of the numerical 
results.  In Table \ref{table:algo} we use $\hat{H}$ to denote a $N\times N$ matrix
and using the fact that a matrix can also be interpreted as a vector, $\vec{H}$ for 
the $N^2$ column vector obtained by concatenating the columns of $\op{H}$ vertically.
$W$ is a $N^2 \times N^2$ matrix whose columns $W_{:,j}$ represent the basis elements
of $L_0$.  Note that $\op{W}_{:,j}$ is the $j$-th basis element interpreted as 
$N \times N$ matrix.

To initialize $W$ we start with $W_{:,1}=\vec{H}_0$ and add $\vec{H}_m$ for $m=1,
\cdots,M$ provided that the additional column increases the rank of $W$.  This 
guarantees that $W$ initially consists of linearly independent generators of $L_0$.  
To construct a basis for $L_0$, we compute all possible commutators of the columns 
$W_{:,j}$ of $W$, interpreted as $N \times N$ matrices.  Whenever a commutator is 
linearly independent of the columns of $W$, we add the commutator as a new column 
to $W$.  Note that if we add a new column to $W$ then we also have to compute the 
commutator of the matrix represented by the new column with all the matrices 
represented by the old columns of $W$.  Hence, we repeat computing the commutators 
of the basis elements represented by $W$ until no new columns have been added in 
the previous step or the rank of $W$ reaches the maximum of $N^2$.
\begin{table}
\begin{tabbing}
\textbf{let} $W=\vec{H}_0$ \\
\textbf{let} $r=\mbox{rank}(W)$ \\
\textbf{for} \= $m=2,\cdots,M+1$ \textbf{do} \\
    \> \textbf{if} \= $\mbox{rank}([W,\vec{H}_m])>r$ \textbf{then} \\
    \>    \> append $W$ by column vector $\vec{H}_m$ \\
    \>    \> $r=r+1$ \\
    \> \textbf{endif} \\
\textbf{endfor} \\
\textbf{let} $r_o=0$ \\
\textbf{let} $r_n=\mbox{rank}(W)$ \\
\textbf{repeat} \\
   \> \textbf{for} $l=r_o+1,\cdots,r_n$ \textbf{do} \\ 
   \> \> \textbf{for} \= $j=1,\cdots,l-1$ \textbf{do} \\
   \> \> \> \textbf{let} $\op{h}=[\op{W}_{:,l}, \op{W}_{:,j}]$ \\
   \> \> \> \textbf{if} \= $\mbox{rank}([W,\vec{h}])>r$ \textbf{then} \\
   \> \> \>    \> append $W$ by column vector $\vec{h}$ \\
   \> \> \>    \> $r=r+1$\\
   \> \> \> \textbf{endif} \\
   \> \> \textbf{endfor} \\
   \> \textbf{endfor}\\
  \> \textbf{let} $r_o=r_n$ \\
  \> \textbf{let} $r_n=\mbox{rank}(W)$ \\
\textbf{until} $r_n=r_o$ \textbf{or} $r=N^2$ 
\end{tabbing}
\caption{Algorithm to compute the Lie algebra generated by a control system $\{\op{H}_0,\cdots,\op{H}_M\}$}
\label{table:algo}
\end{table}

\section{Controllability calculations}
\label{sec:calculations}

We implemented the algorithm presented in the previous section and computed the 
dimension of the dynamical Lie algebra for the system $\op{H}=\op{H}_0+f(t)\op{H}_1$
with $\op{H}_0$ and $\op{H}_1$ as in (\ref{eq:Hzero}) and (\ref{eq:Hone}) for various
choices of the energy levels $E_n$ and the transition dipole moments $d_n$.  
In particular, we studied the $N$-level harmonic oscillator with energy levels
\begin{equation} \label{eq:EHarm}
 E_n = n-\mbox{$\frac{1}{2}$}
\end{equation}
and the $N$-level Morse oscillator with energy levels
\begin{equation} \label{eq:EMorse}
 E_n = (n-\mbox{$\frac{1}{2}$})[1- \mbox{$\frac{1}{2}$} B (n-\mbox{$\frac{1}{2}$})]
\end{equation}
where $B$ is a (usually small) positive real number.  In our numerical computations we 
used $B=0.0419$, which corresponds to a Morse oscillator model of the molecular bond 
for hydrogen fluoride \cite{90CBC}.  We computed the dimension of the Lie algebra $L_0$ 
for systems with varying dimension $N$ and for different choices of the transition 
dipole moments $d_n$.  The results of some of these computations are presented in 
Table \ref{table:dimensions}.

For the Morse oscillator system we observe that the dimension of $L_0$ is always $N^2$,
i.e., it is completely controllable independent of the choice of the $d_n$ (as long as
the $d_n$ are non-zero).

For the harmonic oscillator, however, the dimension of $L_0$ depends on the choice of 
the transition dipole moments.  For the usual choice, $d_n=\sqrt{n}$, the dimension of
$L_0$ is $N^2$, i.e., the system is completely controllable.  However, if we chose all 
the $d_n$ to be equal, e.g., $d_n=1$ for $n=1,\cdots,N-1$, then the dimension of the 
Lie algebra $L_0$ is less than $N^2$ and the system is therefore \emph{not} completely 
controllable for $N>2$.  It is also worth noting that a slight modification of the $d_n$
is sufficient in this case to restore complete controllability: if we choose $d_n=1$,
$n=1,\cdots,N-2$ and $d_{N-1}=2$ then the dimension of $L_0$ is again $N^2$.

The extensive data we gathered strongly suggested that any Morse oscillator system 
with non-zero transition dipole moments, i.e., $d_n\neq 0$, $n=1,\cdots, N-1$, is 
completely controllable for any $N$, while complete controllability for a harmonic 
oscillator seemed to depend on the values of the transition dipole moments $d_n$.  
These observations prompted us to study the issue of controllability systematically
using Lie algebra techniques.  
\begin{table}
\begin{center}
\vspace{4ex}
\begin {tabular}{|l|l||c|c|c|c|c|c|c|}\hline         
         & Dim.\ $N$         & 2 & 3 &  4 &  5 &  6 &  7 &  8 \\\hline\hline
Morse    & $d_n=\sqrt{n}$    & 4 & 9 & 16 & 25 & 36 & 49 & 64 \\\hline
         & $d_n=1$           & 4 & 9 & 16 & 25 & 36 & 49 & 64 \\\hline
         & $d_n=1, d_{N-1}=2$& 4 & 9 & 16 & 25 & 36 & 49 & 64 \\\hline\hline
Harmonic & $d_n=\sqrt{n}$    & 4 & 9 & 16 & 25 & 36 & 49 & 64 \\\hline
         & $d_n=1$           & 4 & 4 & 11 & 11 & 22 & 22 & 37 \\\hline
         & $d_n=1, d_{N-1}=2$& 4 & 9 & 16 & 25 & 36 & 49 & 64 \\\hline
\end{tabular}
\caption{Dimensions of the Lie algebra} \label{table:dimensions}
\end{center}
\end{table}
\section{Results from Lie algebra theory}
\label{sec:theory}
In order to prove our conjectures about complete controllability based on numerical 
evidence, a few general results from the theory of Lie algebras are required.  For 
more detailed information about Lie algebras and Lie groups the reader is referred to
\cite{62Jacobson,97Cornwell,74Gilmore,66Hermann} or any other book on the subject.

We first observe that $u(N)=su(N)\oplus u(1)$, where $su(N)$ is the Lie algebra of
traceless skew-Hermitian $N\times N$ matrices.  If the diagonal elements $d_{m,m}$
of the interaction operators $\op{H}_m$ for $m>0$ are zero, as is the case in the 
dipole approximation, then the interaction operators $\op{H}_m$ are represented by
traceless Hermitian matrices, i.e., $i\op{H}_m\in su(N)$ for $m>0$.  If the internal
Hamiltonian $\op{H}_0$ is traceless as well, i.e., $i\op{H}_0\in su(N)$, then the 
dynamical Lie algebra $L_0$ generated by $i\op{H}_0$ and $i\op{H}_m$ must be $su(N)$,
or a subalgebra of $su(N)$, since the commutator of two traceless skew-Hermitian 
matrices is always a traceless skew-Hermitian matrix.  By our strict definition of
complete controllability, a system whose dynamical Lie algebra is $su(N)$ is \emph{not}
completely controllable since its dynamical Lie group is $SU(N)$, i.e., the Lie group
of unitary $N\times N$ matrices with determinant one, and $SU(N)$ is a proper subgroup
of $U(N)$, the Lie group of all unitary $N \times N$ matrices.  (For a discussion of
the practical significance of the difference between $su(N)$ and $u(N)$ see appendix
\ref{appendix:a}.)  On the other hand, we have the following useful 

\begin{lemma} \label{lemma:one}
If the dynamical Lie algebra $L_0$ contains $su(N)$ and $\op{H}_0$ has non-zero trace
then we have $L_0=su(N) \oplus u(1) \simeq u(N)$.
\end{lemma}

\begin{proof} 
Note that we can write
\begin{equation}
  \op{H}_0 = \frac{1}{N} \trace{\op{H}_0} \op{I} + \op{H}_0',
\end{equation}
where $i\op{H}_0'\in su(N)$.  $i\op{H}_0$ is in $L_0$ by definition.  Since $L_0$ 
contains $su(N)$ and hence $i\op{H}_0'$, it must also contain the identity matrix 
$i\op{I}=(\op{H}_0-\op{H}_0') N/\trace{\op{H}_0}$.  Hence, noting that $\op{I}$ 
generates a one-dimensional Lie algebra isomorphic to $u(1)$ we have indeed 
$L_0=su(N)\oplus u(1)$.
\end{proof}
Thus, in order to show that a system is completely controllable we only need to show
that $\trace{\op{H}_0}\neq 0$ and that $L_0$ contains $su(N)$.  

To verify that $L_0$ contains $su(N)$ we need a complete set of generators for the Lie 
algebra $su(N)$.  Let $\op{e}_{n,n'}$ be the $N\times N$ matrix such that the element 
in the $n$-th row and $n'$-th column is 1 while all other elements are 0, i.e., 
\begin{equation}
 (\op{e}_{n,n'})_{ij}=\delta_{in}\delta_{jn'},
\end{equation} 
where $\delta_{ij}$ is the Kronecker symbol.  One can easily see that any traceless
skew-Hermitian matrix must be a real linear combination of the $N^2-1$ basic matrices
\begin{equation}
\begin{array}{l@{\quad}l} 
  \op{e}_{n,n'}^R = \op{e}_{n,n'} - \op{e}_{n',n}  & 1\le n\le N-1, \; n< n'\le N \\ 
  \op{e}_{n,n'}^I = i(\op{e}_{n,n'}+\op{e}_{n',n}) & 1\le n\le N-1, \; n< n'\le N \\
  \op{h}_n        = \op{e}_{nn}-\op{e}_{n+1,n+1}   & 1\le n\le N-1.
\end{array}
\end{equation}
However, verifying that $L_0$ contains all of the $N^2-1$ basis elements would be quite
tedious.  Fortunately, this is not necessary.

\begin{lemma}
The skew-Hermitian $N\times N$ matrices $\op{e}_{n,n+1}^R$ and $\op{e}_{n,n+1}^I$,
$1\le n< N$, generate the Lie algebra $su(N)$.
\end{lemma}

\begin{proof}
Using the relation
\[
  \op{e}_{n,n'} \op{e}_{m,m'} = \op{e}_{n,m'} \delta_{n'm},
\]
which follows from the definition of $\op{e}_{n,n'}$, it can be verified by direct 
computation that the skew-Hermitian matrices $\op{e}_{n,n'}^R$ and $\op{e}_{n,n'}^I$ 
satisfy the equations
\begin{equation} \label{eq:eij}
 \begin{array}{c}
 {[\op{e}_{n,n'}^R,\op{e}_{n',n''}^R] = \op{e}_{n,n''}^R \quad n''\neq n,} \\
 {[\op{e}_{n,n'}^R,\op{e}_{n',n''}^I] = \op{e}_{n,n''}^I \quad n''\neq n,} \\
 {[\op{e}_{n,n'}^R,\op{e}_{n',n}^I] = 2i(\op{e}_{n,n}-\op{e}_{n',n'})} 
 \end{array}
\end{equation}
and a bit of algebra therefore shows that the $2(N-1)$ elements $\op{e}_{n,n+1}^R$
and $\op{e}_{n,n+1}^I$ for $1\le n \le N-1$ generate the entire Lie algebra $su(N)$.  
\end{proof}

Hence, if $L_0$ contains the generators $\op{e}_{n,n+1}^R$ and $\op{e}_{n,n+1}^I$ for 
$1\le n \le N-1$, then $L_0$ must contain $su(N)$.  Thus, given a system whose energy
levels are well enough separated to permit selective control of each transition
between adjacent energy levels through frequency discrimination, i.e.,
\[
  i\op{H} = i\op{H}_0 + \sum_{n=1}^{N-1} f_n(t) \cos(\mu_n t) i\op{H}_1
\]
with $\op{H}_0$ as in (\ref{eq:Hzero}), $\mu_n=E_n-E_{n+1}$ and
\[
  i\op{H}_n = i d_n (\ket{n}\bra{n+1}+\ket{n+1}\bra{n}) = d_n \op{e}_{n,n+1}^I,
\]
where $f_n(t)$ is slowly time-varying compared to $\cos(\mu_n t)$ and $d_n \neq 0$, 
we can conclude immediately that the Lie algebra of the system contains $su(N)$ and
hence that the system is completely controllable (if $\trace{\op{H}_0} \neq 0$).  To
see this, note that the generators $\op{e}_{n,n+1}^I$ are given and the generators 
$\op{e}_{n,n+1}^R$ can be obtained by computing the commutators
\[
   \mu_n^{-1} [i\op{H}_0,\op{e}_{n,n+1}^I] = \op{e}_{n,n+1}^R. 
\]

However, selective control of individual transitions between adjacent energy levels 
through frequency discrimination is not always possible.  For instance, as pointed 
out earlier, it fails when the energy levels are equally spaced or degenerate, and it 
may not be a good approach for systems with nearly equally spaced energy levels, such
as Morse oscillators.  Furthermore, even if it is possible to use multiple pulses to
selectively control individual transitions, one may not wish to do so.  Instead, one
may for instance wish to control the system with a single optimally shaped control 
pulse obtained using an efficient optimal control algorithm 
\cite{JCP109p385,JCP110p9825,PRA61n012101}

In order to establish criteria for complete controllability of $N$-level systems 
subject to a single control field that drives all permitted transitions concurrently, 
we need another lemma, which makes use of the dipole form of $\op{H}_1$.

\begin{lemma}{} \label{lemma:two}
If $\op{H}_1$ has the special form (\ref{eq:Hone}), i.e.,
\begin{equation}
  i\op{H}_1 =\sum_{n=1}^{N-1} d_n \op{e}_{n,n+1}^I
\end{equation}
then it suffices to show that $L_0$ contains the pair of generators $\op{e}_{12}^R$
and $\op{e}_{12}^I$ or $\op{e}_{N-1,N}^R$ and $\op{e}_{N-1,N}^I$, i.e., if $L_0$
contains either of these two pairs of generators, then it contains all the generators
of $su(N)$.
\end{lemma}

\begin{proof}{}
If $\op{e}_{12}^R, \op{e}_{12}^I\in L_0$ then 
\begin{eqnarray*}
{\op{h}_1}&=&{\frac{1}{2}[\op{e}_{12}^R,\op{e}_{21}^I]
              =i(\op{e}_{11}-\op{e}_{22})\in L_0,}\\
{\op{V}_1}&=&{i\op{H}_1 - d_1 \op{e}_{12}^I
              =\sum_{n=2}^{N-1} d_n \op{e}_{n,n+1}^I\in L_0.}
\end{eqnarray*}
where $\op{e}_{21}^I=\op{e}_{12}^I$.  This leads to 
\begin{eqnarray*}
{[\op{h}_1,\op{V}_1]}             &=& {d_2 \op{e}_{23}^R \in L_0,} \\
{-[\op{h}_1,[\op{h}_1,\op{V}_1]]} &=& {d_2 \op{e}_{23}^I \in L_0.} 
\end{eqnarray*}
Since $d_2 \neq 0$ by hypothesis it follows that $\op{e}_{23}^R,\op{e}_{23}^I\in L_0$.
Repeating this procedure $N-2$ times shows that all the generators $\op{e}_{n,n+1}^R$ 
and $\op{e}_{n,n+1}^I$ for $1\le n<N$ are in $L_0$.  Similarly, we can show that 
$\op{e}_{N-1,N}^R$ and $\op{e}_{N-1,N}^I$ in $L_0$ implies that $L_0$ contains all the
generators $\op{e}_{n,n+1}^R$ and $\op{e}_{n+1,n}^I$, $1\le n<N$.
\end{proof}
Thus, if $\op{H}_1$ has the special form (\ref{eq:Hone}) then it suffices to show that
$\op{e}_{12}^R,\op{e}_{12}^I\in L_0$ in order to conclude that $L_0$ is at least $su(N)$.
\section{Complete controllability for anharmonic systems}
\label{sec:anharmonic}
Let $\mu_n=E_n-E_{n+1}$ for $n=1,\cdots,N-1$ and $\op{V}=i\op{H}_1$.

\begin{theorem} \label{thm:anharmonic}
If $\mu_1 \neq 0$ and $\mu_n^2 \neq \mu_1^2 $ for $n>1$ then the dynamical Lie group
of the system $\op{H}=\op{H}_0+f(t)\op{H}_1$ with $\op{H}_0$ and $\op{H}_1$ as defined
in (\ref{eq:Hzero}) and (\ref{eq:Hone}) is at least $SU(N)$.  If in addition 
$\trace{\op{H}_0}\neq 0$ then the dynamical Lie group is $U(N)$, i.e., the system is 
completely controllable.
\end{theorem}   

\begin{proof}{}
We evaluate
\begin{eqnarray*}
 {[i\op{H}_0,\op{V}]}
     &=&{-\sum_{n=1}^{N-1}\mu_n d_n \op{e}_{n,n+1}^R \equiv -\op{V}',} \\
 {[i\op{H}_0,\op{V}']} 
     &=& {\sum_{n=1}^{N-1}\mu_n^2 d_n \op{e}_{n,n+1}^I\equiv \op{V}''.} 
\end{eqnarray*}
Using $\op{V}$ and $\op{V}''$ we obtain $\op{V}_1\in L_0$, where
\begin{eqnarray*}
 {\op{V}_1} 
     &\equiv& {\op{V}''-\mu_{N-1}^2 \op{V}}\\
     &=&{\sum_{n=1}^{N-2}(\mu_n^2-\mu_{N-1}^2) d_n \op{e}_{n,n+1}^I.}
\end{eqnarray*}
Repeating the previous steps for $i\op{H}_0$ and $\op{V}_1$ leads to $\op{V}_2 \in L_0$,
where
\begin{eqnarray*}
  {\op{V}_2}
     &\equiv& {-[i\op{H}_0, [i\op{H}_0, \op{V}_1]]-\mu_{N-2}^2 \op{V}_1}\\
     &=& {\sum_{n=1}^{N-3}(\mu_n^2-\mu_{N-2}^2)(\mu_n^2-\mu_{N-1}^2)d_n \op{e}_{n,n+1}^I.}
\end{eqnarray*}
After $N-2$ iterations, we have $\op{V}_{N-2}\in L_0$ where
\[
  \op{V}_{N-2}\equiv d_1\left[ \prod_{n=2}^{N-1}(\mu_1^2-\mu_n^2)\right] \op{e}_{12}^I.
\] 
Since by hypothesis $d_1 \prod_{n=2}^{N-1}(\mu_1^2-\mu_n^2) \neq 0$ this means
$\op{V}_{N-2}' = \op{e}_{12}^I \in L_0$ and noting that $\mu_1\neq 0$ we have
\[
  \op{V}_{N-2}''\equiv -\frac{1}{\mu_1} [i\op{H}_0, \op{V}_{N-2}']=\op{e}_{12}^R \in L_0.
\] 
The conclusion now follows from lemmas \ref{lemma:one} and \ref{lemma:two}.
\end{proof}

This theorem shows that for anharmonic systems complete controllability does not depend
on the values of the transition dipole moments $d_n$ (as long as they are non-zero) and
in particular we have the following  
\begin{corollary} \label{cor:Morse}
A quantum system $\op{H}=\op{H}_0+f(t)\op{H}_1$ with $\op{H}_0$ and $\op{H}_1$ as in 
(\ref{eq:Hzero}) and (\ref{eq:Hone}), respectively, and $E_n$ as in (\ref{eq:EMorse}), 
i.e., a Morse oscillator, is completely controllable for arbitrary non-zero values of 
the transition dipole moments $d_n$.
\end{corollary}

It is worth noting that Theorem \ref{thm:anharmonic} also applies to some degenerate 
quantum systems.
\begin{example} \label{ex:degen}
The system $\op{H}=\op{H}_0+f(t)\op{H}_1$ with 
\[
    \op{H}_0=\mbox{diag}\{ E_1, E_2, \cdots, E_2\}, \quad E_1\neq E_2.
\]
and $\op{H}_1$ as in (\ref{eq:Hone}) is completely controllable by theorem 
\ref{thm:anharmonic} despite the fact that energy level $E_2$ has multiplicity
$N-1$.  In fact, the proof of controllability is even simpler for this system 
as we have $\op{V}''=\mu_1^2 d_1\op{e}_{12}^I$ after only one step.
\end{example}

This example begs the question whether the system is still completely controllable if 
we choose 
\[
   \op{H}_0=\mbox{diag}\{ \underbrace{E_1,\cdots, E_1}_{N-1}, E_2\}, \quad E_1\neq E_2
\]
instead. It is obvious that this system does not satisfy the technical condition $\mu_1
\neq 0$.  However, it can easily be shown that this system is completely controllable 
by modifying the proof.  In fact, this example is just a special case of the following 

\begin{theorem}{}
If $\mu_{N-1}\neq 0$ and $\mu_n^2\neq\mu_{N-1}^2 $ for $n<N-1$ then the dynamical Lie 
group of the system $\op{H}=\op{H}_0+f(t)\op{H}_1$ with $\op{H}_0$ and $\op{H}_1$ as 
defined in (\ref{eq:Hzero}) and (\ref{eq:Hone}), respectively, is at least $SU(N)$.  
If in addition $\mbox{Tr}(\op{H}_0)\neq 0$ then the dynamical Lie group is $U(N)$, 
i.e., the system is completely controllable.
\end{theorem} 
The proof of this theorem is analogous to the proof of Theorem \ref{thm:anharmonic}.
\section{Complete controllability for harmonic systems}
\label{sec:harmonic}
The condition $\mu_1^2\neq \mu_n^2$ in the previous theorems excludes any system with
equally spaced energy levels, such as a harmonic oscillator, for which we have $\mu_1
=\mu_2=\cdots=\mu_{N-1}$.  We shall assume that $\mu_1\neq0$ since for $\mu_1=0$ the 
system is completely degenerate with only one energy level.  

In order to state our theorem, we need to introduce some technical parameters 
\begin{equation} \label{condition2}
     v_n=\left\{ \begin{array}{ll}
                      2 d_1^2-d_2^2,  & n=1;\\
                      2d_n^2-d_{n-1}^2-d_{n+1}^2, & n=2,\cdots,N-2; \\
                      2d_{N-1}^2-d_{N-2}^2,   & n=N-1. 
                    \end{array}  \right.   
\end{equation}
Observe that these technical parameters depend on the values of the transition dipole
moments $d_n$.

\begin{theorem}  \label{thm:harmonic}
The dynamical Lie group for a quantum system $\op{H}=\op{H}_0+f(t)\op{H}_1$ with 
$\op{H}_0$ and $\op{H}_1$ as in (\ref{eq:Hzero}) and (\ref{eq:Hone}) and $N$ equally 
spaced energy levels  
\[ 
   E_n=E_1+(n-1)\mu_1, \quad n=1,\cdots,N, \; \mu_1\neq 0,
\] 
is at least $SU(N)$ if the the parameters $v_n$ satisfy one of the following conditions
\begin{enumerate}
\item $v_n \neq v_{N-1}$ for $1\le n \le N-2$;
\item $v_n \neq v_1$ for $2\le n\le N-1$.
\end{enumerate}
If in addition $\mbox{Tr}(\op{H}_0)\neq 0$ then the dynamical Lie group is $U(N)$, i.e.,
the system is completely  controllable.
\end{theorem}

\begin{proof}
Let $\op{V}=i\op{H}_1$.  In this case the element 
\[
   \tilde{V}=-\mu_1^{-1}[i\op{H}_0,\op{V}]= \sum_{n=1}^{N-1} d_n \op{e}_{n,n+1}^R
\]   
is in $L_0$ and its sum and difference with $\op{V}$ give rise to 
\begin{eqnarray*}
    {\op{V}_1^+} &\equiv &{\sum_{n=1}^{N-1} d_n 
     \underbrace{(\op{e}_{n,n+1}^I+\op{e}_{n,n+1}^R)}_{\equiv \op{e}_{n,n+1}^+},} \\
    {\op{V}_1^-} &\equiv &{\sum_{n=1}^{N-1} d_n 
     \underbrace{(\op{e}_{n,n+1}^I-\op{e}_{n,n+1}^R)}_{\equiv \op{e}_{n,n+1}^-}}
\end{eqnarray*}
which, along with their commutator
\begin{eqnarray*}
   \op{V}_1^0 &=& {\frac{1}{4} [\op{V}_1^+, \op{V}_1^-]}\\
              &=& {id_1^2   \op{e}_{1,1}+\sum_{n=2}^{N-1}i(d_n^2-d_{n-1}^2)\op{e}_{n,n}
                  -id_{N-1}^2 \op{e}_{N,N},}
\end{eqnarray*}
are in $L_0$. Starting with $\op{V}_1^0$ and $\op{V}_1^+$, we have
\[
 [\op{V}_1^0, \op{V}_1^+]=\sum_{n=1}^{N-1} v_n d_n \op{e}_{n,n+1}^+ \in L_0,
\]
\begin{eqnarray*}
 {\op{V}^+_2}
 &\equiv& {[\op{V}_1^0, \op{V}_1^+]-v_1 \op{V}^+_1}\\
 &=& {\sum_{n=2}^{N-1} (v_n-v_1) d_n \op{e}_{n,n+1}^+ \in L_0,}\\
 {\op{V}^+_3}
 &\equiv& {[\op{V}_1^0, \op{V}^+_2]-v_2\op{V}^+_2}\\
 &=& {\sum_{n=3}^{N-1}(v_n-v_2)(v_n-v_1) d_n \op{e}_{n,n+1}^+ \in L_0,}\\
 &\vdots &\\         
 {\op{V}^+_{n-1}}
 &\equiv& {[\op{V}_1^0, \op{V}^+_{N-2}]-v_{N-2}\op{V}^+_{N-2}}\\
 &=& {d_{N-1}\left[ \prod_{n=1}^{N-2} (v_{N-1}-v_n)\right] \op{e}_{N-1,N}^+ \in L_0.}
\end{eqnarray*}
Since by hypothesis $v_{N-1} \neq v_n$ for $n=1,2,\cdots,N-2$, we have 
$\op{e}_{N-1,N}^+\in L_0$. 

Similarly, starting with $\op{V}^0$ and $\op{V}_1^-$, we can also prove 
$\op{e}_{N-1,N}^- \in L_0$.  This implies $\op{e}_{N-1,N}^R$ and $\op{e}_{N-1,N}^I$
are in $L_0$ and hence the conclusion follows from lemmas \ref{lemma:one} and
\ref{lemma:two}.
\end{proof}

\begin{example} 
The standard $N$-level harmonic oscillator system $\op{H}=\op{H}_0+f(t)\op{H}_1$ with 
$\op{H}_0$ and $\op{H}_1$ as in (\ref{eq:Hzero}) and (\ref{eq:Hone}), $E_n=n-1/2$ 
and $d_n=\sqrt{n}$ is completely controllable by theorem \ref{thm:harmonic} since
we have
\[
    v_1=v_2=\cdots =v_{N-2}=0, \quad v_{N-1} = N,
\]
i.e., the $v_n$ satisfy $v_{N-1}-v_n=N$ for $n=1,\cdots,N-2$.
\end{example}

\begin{example}
However, the $N$-level harmonic oscillator system $\op{H}=\op{H}_0+f(t)\op{H}_1$ with 
$\op{H}_0$ and $\op{H}_1$ as in (\ref{eq:Hzero}) and (\ref{eq:Hone}), $E_n=n-1/2$ and 
$d_n=1$ for $n=1,\cdots, N-1$ does \emph{not} satisfy the hypothesis of theorem 
\ref{thm:harmonic} since we have
\[
    v_1=1, \quad v_2=\cdots =v_{N-2}=0, \quad v_{N-1}=1,
\]
i.e., $v_{N-1}=v_1$.  Numerical calculations for various $N$ confirm that this system
is indeed not completely controllable. (See Table \ref{table:dimensions}.)  

For $N=4$ one can, e.g., verify either numerically or by analyzing the Lie algebra that
the simple skew-Hermitian matrix
\[
  \op{e}_{12}^I = \left( \begin{array}{cccc}
                   0 & i & 0 & 0 \\
                   i & 0 & 0 & 0 \\
                   0 & 0 & 0 & 0 \\
                   0 & 0 & 0 & 0
                   \end{array} \right)
\]  
is not in $L_0$ and therefore the unitary operator 
\[
  \op{U}(\theta)=\exp(\theta\op{e}_{12}^I) = \left( \begin{array}{cccc}
                   \cos(\theta) & i\sin(\theta)& 0 & 0 \\
                   i\sin(\theta)& \cos(\theta) & 0 & 0 \\
                   0       & 0       & 1 & 0 \\
                   0       & 0       & 0 & 1
                   \end{array} \right)
\]  
can not be dynamically generated for this system for any $\theta\in (0,2\pi)$.  Thus, if 
the system is initially in state $\op{\rho}_0=\sum_{n=1}^4 w_n \ket{n}\bra{n}$ then the 
target state 
\[
 \op{\rho}(t_F)=\op{U}(\theta)\op{\rho}_0\op{U}(\theta)^\dagger
\]
is generally not dynamically accessible since it is impossible to put the energy
eigenstates $\ket{1}$ and $\ket{2}$ into superposition without equally ``entangling''
the states $\ket{3}$ and $\ket{4}$ of the initial ensemble.
\end{example}

\begin{example}
The $N$-level harmonic oscillator system $\op{H}=\op{H}_0+f(t)\op{H}_1$ with $\op{H}_0$
and $\op{H}_1$ as in (\ref{eq:Hzero}) and (\ref{eq:Hone}), $E_n=n-1/2$, $d_n=1$ for 
$n=1,\cdots, N-2$ but $d_{N-1}=2$, however, \emph{does} satisfy the hypothesis of 
theorem \ref{thm:harmonic} since we have
\[
    v_1=1, \; v_2=\cdots =v_{N-3}=0, \; v_{N-2}=-3, \; v_{N-1}=7,
\]
i.e., $v_n\neq v_1$ for $n\neq 1$.  Numerical calculations for various $N$ confirm that
this system is completely controllable (see Table \ref{table:dimensions}).  
\end{example}

The rather surprising results of the previous two examples can be understood by 
analyzing the Lie algebra generated by $\op{H}_0$ and $\op{H}_1$ in both cases.  
Although the details of this analysis are beyond the scope of this paper (and 
will be discussed in a future paper) we would like to mention here that for a 
harmonic system with transition dipole moments satisfying the ``symmetry'' 
relation $d_{N-n}=\pm d_n$ for $1\le n \le N/2$, the transitions $n\rightarrow 
n+1$ and $N-n\rightarrow N+1-n$ for $1\le n \le N/2$ become coupled.  This leads
to a collapse of the Lie algebra and loss of complete controllability, which can
be restored by breaking the symmetry in the transition dipole moments.  This is 
why changing $d_{N-1}=1$ to $d_{N-1}=2$ restored controllability in the last 
example.  In fact, changing $d_{N-1}$ to any value other than $\pm 1$ would work
as well.
\section{Conclusion}
\label{sec:conclusion}

The question of complete controllability of quantum systems using external control 
fields has been addressed before by various authors and it is, for instance, well 
known that a quantum system is completely controllable if it is possible to address
a sufficiently large set of single transitions using multiple frequency-selective 
control pulses.  However, many optimization strategies attempt to find a single 
control pulse that addresses all transitions concurrently to achieve the control 
objective.  Furthermore, control based on frequency discrimination is not always 
possible, e.g., it is not suitable for systems with equally or almost equally
spaced or degenerate energy levels.  

Despite the relative importance of control strategies involving a single control 
pulse, sufficient criteria for complete controllability in this case have so far
been missing.  In this paper we addressed this problem and established general 
criteria for complete controllability of quantum systems subject to a single 
control pulse.  

In particular, we showed that most anharmonic, non-decomposable quantum systems 
are completely controllable using a single control that drives all the transitions
concurrently, independent of the values of the transition dipole moments $d_n$.  
For quantum systems with equally spaced energy levels we demonstrated that complete 
controllability depends on the values of the transition dipole moments $d_n$ and 
derived conditions that guarantee complete controllability.  

We verified that the standard truncated harmonic oscillator with transition dipole 
moments $d_n=\sqrt{n}$ satisfies these conditions and gave examples of harmonic 
systems that do not satisfy the conditions.  In the latter case we also checked by
direct computation of the Lie algebra that they are not completely controllable 
and showed that there are certain unitary operators that can not be dynamically
realized for these systems.
\section{Acknowledgements}
We would like to thank F.~C.\ Langbein and B.~I.\ Mills of Cardiff University and 
J.~V.\ Leahy of the University of Oregon for providing valuable comments and suggestions.

\appendix
\section{Controllability and the question of trace}
\label{appendix:a}
All the theorems about complete controllability presented in this paper require 
that the trace of $\op{H}_0$ be non-zero.  The mathematical necessity of this 
hypothesis is obvious: a set of traceless skew-Hermitian $N\times N$ matrices 
cannot generate all of $u(N)$ but at most $su(N)$.  Hence, assuming that the 
interaction terms $\op{H}_m$, $m>0$, are represented by traceless skew-Hermitian
matrices, if $\op{H}_0$ is traceless as well then the dynamical Lie group can be
at most $SU(N)$, the set of unitary matrices with determinant one, which is a 
proper subgroup of $U(N)$.  However, our definition of complete controllability 
requires that all unitary matrices be dynamically accessible.  

Nevertheless, the trace condition is physically somewhat disturbing since the 
energy levels of a physical system are generally only determined up to a constant,
and hence the trace of $\op{H}_0$ seems physically rather insignificant as one can
always make it either zero or non-zero by shifting the energy levels of the system
by a constant.  We shall attempt to resolve this apparent conflict by showing that
the difference between $SU(N)$ and $U(N)$ is only a phase factor.  

Let the initial state of the system be represented by the normalized wavefunction 
$\ket{\psi_0}$.  If the dynamical Lie group of the system is $U(N)$ then any other
pure state represented by normalized wavefunction $\ket{\psi_1}$ is dynamically 
reachable since given any two normalized wavefunctions there always exists a (not 
necessarily unique) unitary transformation $\op{U}$ such that $\ket{\psi_1}=\op{U}
\ket{\psi_0}$ and we can find a path $\gamma(t)=\op{U}(t,t_0)$ in $U(N)$ such that
$\gamma(t_0)=\op{1}$ and $\gamma(t_F)=\op{U}$.  Since the determinant of a unitary
operator is a complex number of modulus 1, we can write $\det(\gamma(t))=e^{i\phi(t)}$. 
Noting that $\det(\alpha A)=\alpha^N \det(A)$ where $N$ is the size of $A$, we see
immediately that $\tilde{\gamma(t)}\equiv e^{-i\phi(t)/N}\gamma(t)$ has determinant 1
\[
  \det(\tilde{\gamma}(t)) = (e^{-i\phi(t)/N})^N \det(\gamma(t))
                          = e^{-i\phi(t)} e^{i\phi(t)} \equiv 1,
\]
and thus defines a path in $SU(N)$.  Furthermore,
\[
  \tilde{\gamma}(t_F) \ket{\psi_0} = e^{-i\phi/N} \op{U} \ket{\psi_0} 
                                   = e^{-i\phi/N} \ket{\psi_1},
\]
i.e., $\tilde{\gamma}(t_F)\ket{\psi_0}$ and $\ket{\psi_1}$ differ only by a phase 
factor.  Hence, if the dynamical Lie group is $SU(N)$ then we loose control over 
the phase of the state, otherwise there is no difference.  

Thus, for practical applications that do not require phase control one need not 
worry about the trace.  For instance, if the goal of controlling the system is to 
maximize the expectation value of an observable $\op{A}$ at a target time $t_F$, 
\[
  \langle\op{A}(t_F) \rangle = \bra{\psi(t_F)} \op{A} \ket{\psi(t_F)},
\]
then clearly the phases of the target states are irrelevant as they are cancelled
out by computing the expectation value anyway.

Moreover, if the initial state is given by a density matrix  $\op{\rho}_0$ then 
any target state $\op{\rho}(t_F)$ that is dynamically accessible via a path in 
$U(N)$ is also dynamically accessible via a path in $SU(N)$.  To see this, let 
$\gamma(t)=\op{U}(t,t_0)$ be a path in $U(N)$ such that 
\begin{equation}
  \op{\rho}(t_F) = \op{U}(t_F,t_0) \op{\rho}_0 \op{U}(t_F,t_0)^\dagger.
\end{equation}
Again, we have $\det(\gamma(t))=e^{i\phi(t)}$ and $\tilde{\gamma}(t)=e^{-i\phi(t)/N}
\gamma(t)$ defines a path in $SU(N)$ that is equivalent to $\gamma(t)$ since 
\footnote{Note that if $A$ is unitary then $\det(A^\dagger)=[\det(A)]^{-1}$, i.e., 
$\det(\gamma(t)^\dagger)=e^{-i\phi(t)}$.} 
\begin{eqnarray*}
 \op{\rho}(t) 
 &=& \gamma(t) \op{\rho}_0 \gamma(t)^\dagger \\
 &=& e^{i\phi(t)/N}\tilde{\gamma}(t)\op{\rho}_0\tilde{\gamma}(t)^\dagger e^{-i\phi(t)/N}\\
 &=& \tilde{\gamma}(t) \op{\rho}_0 \tilde{\gamma}(t)^\dagger,
\end{eqnarray*}
i.e., the phase factors $e^{-i\phi(t)/N}$ cancel out completely.  

However, there are some applications of control in quantum computation where it is 
important to have phase control and when $SU(N)$ is not adequate.  Therefore, we have
chosen to require the dynamical Lie group to be $U(N)$ for complete controllability.

\bibliography{books,papers2000,papers9599,science}
\end{document}